\documentclass[
reprint,
superscriptaddress,
amsmath,
amssymb,
showkeys,
aps,
prb,
nolongbibliography,
]{revtex4-2}

\usepackage{threeparttable}
\usepackage{setspace}
\usepackage{makecell}
\usepackage{amsmath}
\usepackage{graphicx}
\usepackage{dcolumn}
\usepackage{bm}
\usepackage{hyperref}
\usepackage[mathlines, pagewise]{lineno}
\usepackage{textcomp}
\usepackage{gensymb}
\usepackage{subfigure}
\usepackage{url}
\usepackage{xcolor}
\usepackage{soul}
\usepackage{hhline}
\usepackage[version=3]{mhchem}
\usepackage{multirow}
\usepackage[american]{babel}
\usepackage{textgreek}
\usepackage{miller}

\begin{document}

\title{
Phase glides and self-organization of atomically abrupt interfaces out of stochastic disorder in \texorpdfstring{$\alpha$-\ce{Ga2O3}}{}
}

\author{Alexander Azarov} \email{alexander.azarov@smn.uio.no}
\affiliation{Department of Physics and Centre for Materials Science and Nanotechnology, University of Oslo, PO Box 1048 Blindern, N-0316 Oslo, Norway}

\author{Javier Garc{\'i}a Fern{\'a}ndez}
\affiliation{Department of Physics and Centre for Materials Science and Nanotechnology, University of Oslo, PO Box 1048 Blindern, N-0316 Oslo, Norway}

\author{Junlei Zhao} \email{zhaojl@sustech.edu.cn}
\affiliation{Department of Electronic and Electrical Engineering, Southern University of Science and Technology, Shenzhen 518055, China}

\author{Ru He}
\affiliation{Department of Physics and Helsinki Institute of Physics, University of Helsinki, P.O. Box 43, FI-00014, Finland}

\author{Ji-Hyeon Park}
\affiliation{Korea Institute of Ceramic Engineering \& Technology, Jinju 52851, South Korea}

\author{Dae-Woo Jeon}
\affiliation{Korea Institute of Ceramic Engineering \& Technology, Jinju 52851, South Korea}

\author{{\O}ystein Prytz}
\affiliation{Department of Physics and Centre for Materials Science and Nanotechnology, University of Oslo, PO Box 1048 Blindern, N-0316 Oslo, Norway}

\author{Flyura Djurabekova} 
\affiliation{Department of Physics and Helsinki Institute of Physics, University of Helsinki, P.O. Box 43, FI-00014, Finland}

\author{Andrej Kuznetsov} \email{andrej.kuznetsov@fys.uio.no}
\affiliation{Department of Physics and Centre for Materials Science and Nanotechnology, University of Oslo, PO Box 1048 Blindern, N-0316 Oslo, Norway}

\begin{abstract} 

Disorder-induced ordering and unprecedentedly high radiation tolerance in $\gamma$-phase of gallium oxide is a recent spectacular discovery at the intersection of the fundamental physics and electronic applications. 
Importantly, by far, these data were collected with initial samples in form of the thermodynamically stable $\beta$-phase of this material. 
Here, we investigate these phenomena starting instead from already metastable $\alpha$-phase and explain radically new trend occurring in the system. 
We argue that in contrast to that in $\beta$-to-$\gamma$ disorder-induced transitions, the O sublattice in $\alpha$-phase exhibits hexagonal close-packed structure, so that to activate $\alpha$-to-$\gamma$ transformation significant structural rearrangements are required in both Ga and O sublattices. 
Moreover, consistently with theoretical predictions, $\alpha$-to-$\gamma$ phase transformation requires accumulation of the substantial tensile strain to initiate otherwise impossible lattice glides. 
Thus, we explain the experimentally observed trends in term of the combination of disorder and strain governing the process. 
Finally, and perhaps most amazingly, we demonstrate atomically abrupt $\alpha$/$\gamma$ interfaces paradoxically self-organized out of the stochastic disorder.

\end{abstract}

\maketitle

\section*{Introduction}

Recently gallium oxide (\ce{Ga2O3}) has attracted attention of a broad audience spreading from those dealing with fundamentals of the phase transitions [1-5] to device application experts [6-10]. 
Among the rest of the highlights there was a discovery of disorder-induced ordering in \ce{Ga2O3} [11-14] and unprecedently high radiation tolerance of the formed structures [15]. 
Specifically, it was shown that even though its thermodynamically stable monoclinic polymorph ($\beta$-\ce{Ga2O3}) can be swiftly disordered, it does not amorphize under irradiation, but converts to a cubic defective spinel polymorph ($\gamma$-\ce{Ga2O3}), remaining crystalline independently of subsequent irradiation [15]. 
Moreover, electronic radiation tolerance tests performed by comparing Schottky diodes fabricated out of $\beta$- and $\gamma$-polymorphs showed that the $\gamma$-\ce{Ga2O3}-based diodes remain functional, while $\beta$-\ce{Ga2O3}-based diodes lost their rectification under identical irradiation conditions [16]. 
As explained recently, the rationale behind this remarkable $\beta$-to-$\gamma$ \ce{Ga2O3} polymorph transformation is because the oxygen sublattice in these polymorphs, exhibiting face-centered cubic (\textit{fcc}) structure, demonstrates strong recrystallization trends, while the Ga sublattice is susceptible to disorder [17,18]. 
Very recently, this idea was exploited to demonstrate multiple $\gamma$/$\beta$ polymorph repetitions by adjusting spatial distributions of the disorder levels as a function of the irradiation temperature and ion flux [19]; as such demonstrating ``polymorph heterostructures'' not being realized by conventional growth methods otherwise.

Meanwhile, understanding of the radiation phenomena in other \ce{Ga2O3} polymorphs is much less mature. 
For instance, for the metastable rhombohedral polymorph ($\alpha$-\ce{Ga2O3}) there are only a few studies devoted to the radiation defect formation [20-22]; however, indicating that $\alpha$-phase is more radiation resistant at the range of the nuclear stopping power maximum as compared to that of $\beta$-\ce{Ga2O3} [20]. 
Concurrently, the disorder buildup in $\alpha$-\ce{Ga2O3} involves surface amorphization, somewhat resembling the features observed in GaN [23,24]. Nevertheless, if disorder-induced polymorphism is realized in $\alpha$-\ce{Ga2O3} its impact on the device applicability may be even more interesting than that in $\beta$-\ce{Ga2O3}; since $\alpha$-\ce{Ga2O3} exhibits the widest bandgap among the rest of the \ce{Ga2O3} polymorphs family [25,26]; making it more likely to anticipate higher band offsets in, \textit{e.g.}, $\gamma$/$\alpha$ interfaces [27]. 

Thus, in the present work we undertook a systematic investigation of the radiation phenomena in $\alpha$-\ce{Ga2O3} and determined conditions sufficient for igniting $\alpha$-to-$\gamma$ polymorph transition. 
We argued that in contrast to that in $\beta$-phase, the O sublattice in $\alpha$-phase posses hexagonal close-packed (\textit{hcp}) structure, so that to activate $\alpha$-to-$\gamma$ phase transformation significant structural rearrangements are required in both Ga and O sublattices. 
Moreover, consistently with predictions from the energy diagram, $\alpha$-to-$\gamma$ phase transformation requires accumulation of the substantial tensile strain to initiate otherwise impossible lattice glides. 
Thus, we explain these fascinating phase transformation trends in term of the combination of disorder and strain governing the process.
As a result, we demonstrate atomically abrupt $\alpha$/$\gamma$ interfaces paradoxically self-organized out of the stochastic disorder.

\section*{Results and discussion}

\begin{figure*}[ht!]
 \includegraphics[width=16cm]{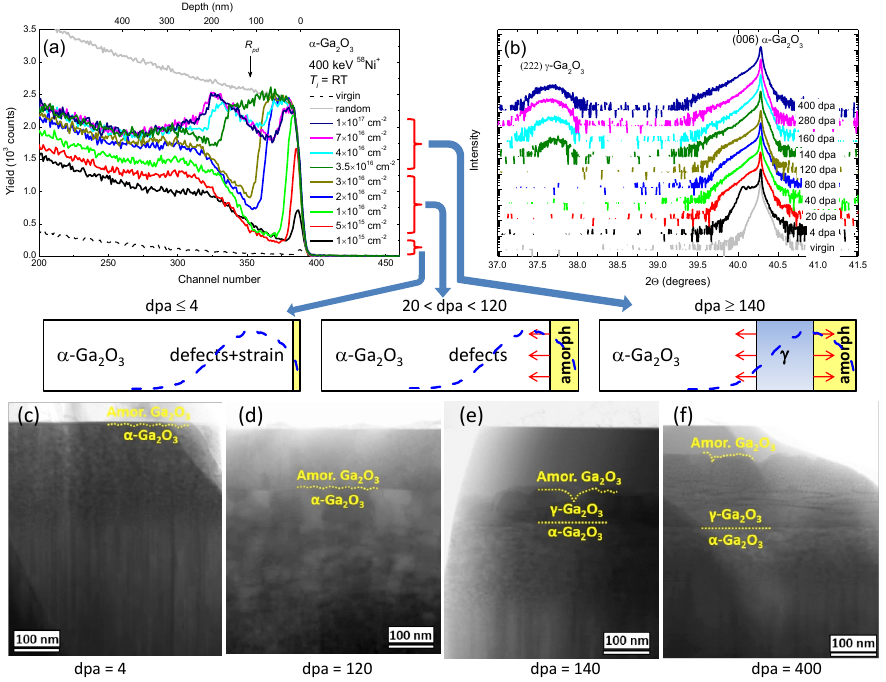}
 \caption{
    \textbf{Phase transformations in $\alpha$-\ce{Ga2O3} as a function of dpa.} 
    (a) RBS/C spectra and (b) corresponding XRD 2$\Theta$ scans of the $\alpha$-\ce{Ga2O3} samples irradiated with 400 keV Ni ions up to the doses as dpa values indicated in legends in the panels (a) and (b), respectively. 
    Panels (c)-(f) show low magnification BF-STEM cross-sections of the selected samples for dpa values highlighting characteristic trends, in correlation with cartoons in the middle of the figure, where the primary defect generation profiles are shown by the dashed lines. 
 }
 \label{fig:1}
\end{figure*}

Fig.~\ref{fig:1} provides a survey of the experimental data including systematic measurements of the samples crystallinity as a function of displacement per atom (dpa) obtained by (a) RBS/C and (b) XRD in combination with TEM cross-sections of the selected samples in panels (c)-(f), associated with characteristic process stages as illustrated in cartoon insets in the middle of the figure. 
Indeed, already for low dpa conditions, \textit{i.e.}, $\leq 4$ dpa, the RBS/C spectra reveal -- consistently with the literature~[20,21] – a surface disorder peak (see Fig.~1(a)), specifically at $\mathrm{dpa}=4$ corresponding to a 6 nm thick amorphous layer, as confirmed by TEM data in Fig.~\ref{fig:1}(c). 
In addition, there is a broader ``bulk'' disorder peak localized far beyond of the maximum of the primary defect generation ($R_\mathrm{pd} \simeq 105$ nm) according to the SRIM code~[28] simulations. 
This stage is accompanied by a tensile strain accumulation as clearly seen from appearing of a shoulder on the left-hand side of the $\alpha$-\ce{Ga2O3} \hkl(006) reflection in the XRD 2$\Theta$ scans (Fig.~\ref{fig:1}(b)) for $\mathrm{dpa} = 4$. 
Further, at $20 \leq \mathrm{dpa} \leq 120$ range, the surface disorder peak broadens and eventually reaching the random level, while the magnitude of the bulk peak saturates at much lower disorder level, see Fig.~\ref{fig:1}(a). 
For example, at $\mathrm{dpa} = 120$, $\sim 130$ nm thick amorphous layer is revealed by TEM as illustrated in Fig.~\ref{fig:1}(d). 
Notably, this disorder accumulation stage does not reveal much of changes in the XRD spectra, except of the strain release, as seen from the evolution of the left-hand side parts of the \hkl(006) diffraction peak in Fig.~\ref{fig:1}(b). 

Spectacularly, additional relatively tiny dpa increase -- just by a few tens of percents -- dramatically changes the structure. 
Indeed, at $\mathrm{dpa} = 140$ RBS/C intensity increases right behind the surface amorphous layer, see the 100-170 nm range below the surface in Fig.~\ref{fig:1}(a). 
This prominent transformation is accompanied by an appearance of a new diffraction peak centered at $\sim 37.7\degree$, see Fig.~\ref{fig:1}(b), which is identified as $\gamma$-\ce{Ga2O3} \hkl(222) reflection [29], in agreement with TEM data in Fig.~\ref{fig:1}(e).
Further dpa increase improves the crystallinity in this region as clearly seen from the decreased the RBS/C yield at $\mathrm{dpa} = 160$ as compared to that at $\mathrm{dpa} = 140$, see Fig.~\ref{fig:1}(a). 
Simultaneously, the width of the phase-modified layer expands with increasing dpa. 
Notably, the $\gamma$-layer expands both into the crystal bulk in the form of $\alpha$-to-$\gamma$ transition, and towards the surface, so that the amorphous layer converts into $\gamma$-phase too, as schematically shown in the corresponding cartoon inset. 
Thus, the surface amorphous layer broadens as a function of dpa until the $\alpha$-to-$\gamma$ phase transformation starts, while further dpa increase leads to the shrinkage of the amorphous layer due to the $\gamma$-film expansion. 
Notably, dpa increase beyond 140 dpa has practically no impact on the crystallinity of the newly formed $\gamma$-phase confirming its remarkable radiation tolerance consistently with literature [15]. 
Moreover, comparing it to $\beta$-phase, $\alpha$-\ce{Ga2O3} itself can be indeed classified as a higher radiation tolerant material, since $\alpha$-to-$\gamma$ phase transition starts at much higher dpa levels (N.B. $\mathrm{dpa} = 1$ was shown to be sufficient to start $\beta$-to-$\gamma$ transition [15]).

\begin{figure*}[ht!]
 \includegraphics[width=16cm]{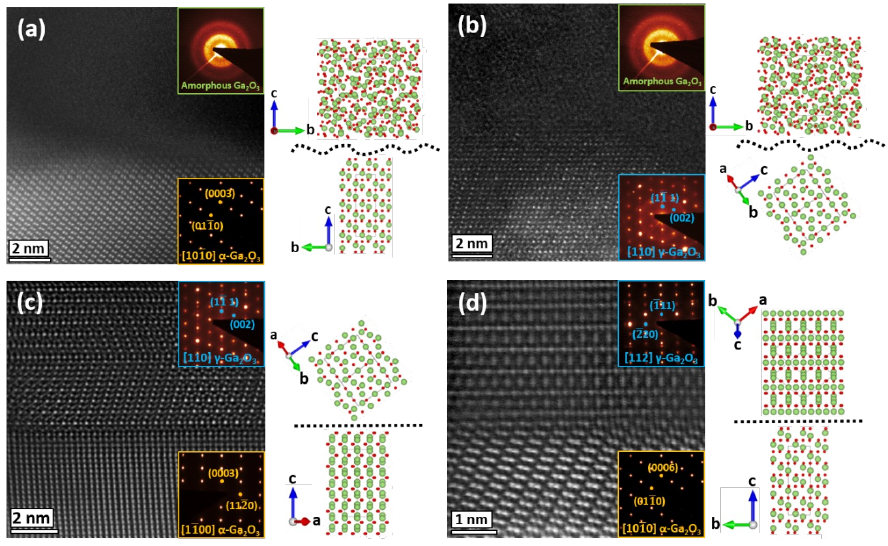}
 \caption{
    \textbf{HRTEM images showing quality of interfaces formed out of stochastic disorder.} 
    (a) Non-abrupt amorphous/$\alpha$ and (b) amorphous/$\gamma$ interfaces in the samples with 4 and 400 dpa, respectively, contra atomically sharp $\gamma$/$\alpha$ interface (imaged along the (c) \hkl[110] $\gamma$-\ce{Ga2O3} / \hkl[1-100] $\alpha$-\ce{Ga2O3} and (d) \hkl[112] $\gamma$-\ce{Ga2O3} / \hkl[10-10] $\alpha$-\ce{Ga2O3} zone axes for the sample with $\mathrm{dpa} = 400$.
    On the right-hand sides of the images is depicted the projected structure model for each phase. 
    Color code: Ga in green, O in red. The interfaces in the insets are shown by the dashed black lines. 
 }
 \label{fig:2}
\end{figure*}

Meanwhile, another spectacular observation indicated already by the data in Figs~\ref{fig:1}(e) and \ref{fig:1}(f) is an ultimate abruptness of the $\gamma$/$\alpha$ interfaces resulted out of colossal disordering process having stochastic nature. 
To investigate this phenomenon in details we performed HAADF-STEM analysis of the interfaces obtained in this study, see Fig.~\ref{fig:2}. 
Specifically, Figs.~\ref{fig:2}(a) and \ref{fig:2}(b) show the high resolution images of the amorphous/$\alpha$-phase and amorphous/$\gamma$-phase interfaces formed in the samples upon disordering with $\mathrm{dpa} = 4$ and 400, respectively. 
As expected from the stochastic nature of disorder, these interfaces are not abrupt, and in case of the amorphous/$\gamma$-phase interface even rather rough. 
However, even it is contraintuitive, polymorph interfaces formed out of the same stochastic disorder, specifically $\gamma$/$\alpha$ interface is atomically sharp as clearly demonstrated by Figs.~\ref{fig:2}(c) and \ref{fig:2}(d) showing high resolution HAADF-STEM images of the interface region taken along different zone axes in the sample subjected to $\mathrm{dpa} = 400$. 
The corresponding fast Fourier transformation (FFT) for each image and the schematic unit cell and lattice stackings oriented as in the interfaces are shown in the insets and at the right-hand sides of the corresponding images, respectively. 
Specifically, the orientation relationships at the $\gamma$/$\alpha$ interface were determined from STEM as $\gamma$\hkl[110]/$\alpha$\hkl[1-100] and $\gamma$\hkl[112]/$\alpha$\hkl[10-10] and were used further in simulations to shed more light on the mechanism of the $\gamma$/$\alpha$ interface formation.   

\begin{figure*}[ht!]
 \includegraphics[width=16cm]{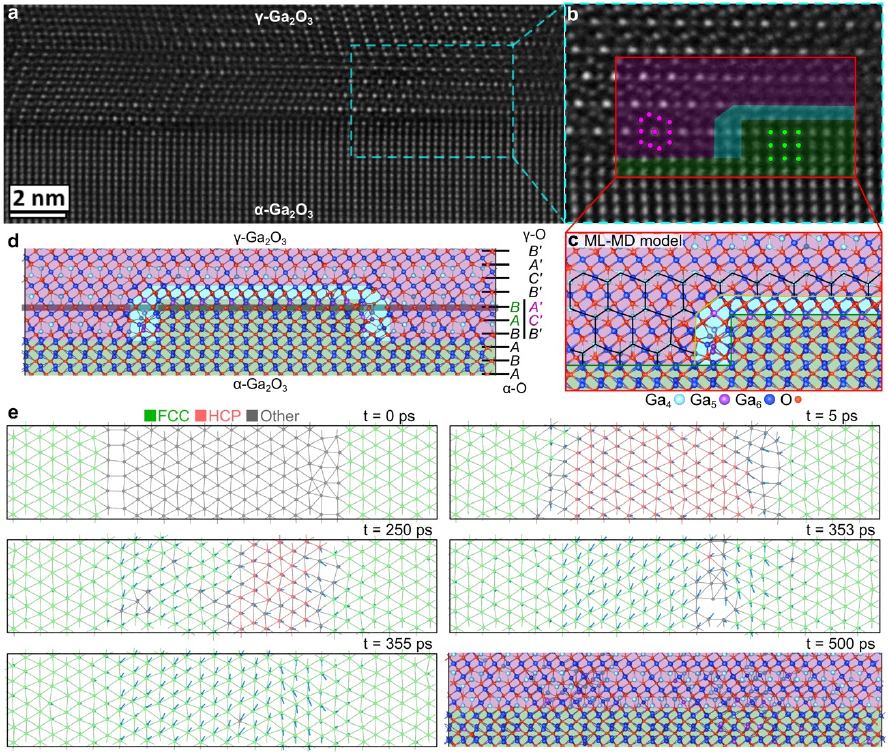}
 \caption{
    \textbf{Atomic configuration and dynamic evolution towards atomically sharp $\gamma$/$\alpha$ interface.} 
    (a) Experimentally measured $\gamma$/$\alpha$ interface as that in Fig.~2(c); 
    (b) Enlarged interface image with the simulation template inset illustrating corresponding initial configurations of the ML-MD model in panel (c); 
    (d) detailed configuration in terms of the O sublattice repetitions in $\gamma$-\ce{Ga2O3} ($\gamma$-O) and $\alpha$-\ce{Ga2O3} ($\alpha$-O); Ga atoms are big blue (colored based on coordination numbers) and O atoms are small red spheres. 
    The $\gamma$-phase, boundary, and $\alpha$-phase regions are colored in purple, cyan and green, respectively. 
    (e) Dynamic evolution of the initially mismatched oxygen ($A’ \vert B$) stacking layer [labeled by gray shadow in panel (d)] at 900 K, viewed from $\alpha$\hkl[000-1]/$\gamma$\hkl[11-1] axis, demonstrating the planar slip displacements relative to the initial positions (indicated by blue arrows). 
    The color-coding of oxygen atoms illustrates local stacking types, such as \textit{fcc} and \textit{hcp}. 
    At $t = 500$ ps, the atomically sharp $\gamma$/$\alpha$ interface forms with a perfectly matched oxygen ($B = B’$) stacking layer in accordance with the ML-MD simulation. 
    See Supplementary Video I for the whole evolution process.
 }
 \label{fig:3}
\end{figure*}

Thus, for that matter, we performed machine-learning-driven molecular dynamics (ML-MD) simulations and Fig.~\ref{fig:3} summarizes the atomic configurations and dynamic evolution of such interface. 
The initial local atomic configuration of the ML-MD interface (Fig.~\ref{fig:3}(c)) closely resembles the interface observed in high-resolution STEM images (Figs.~\ref{fig:3}(a) and \ref{fig:3}(b)). The $\gamma$-O and $\alpha$-O sublattices follow \textit{fcc} ($A’B’C’$-$A’B’C’\dots$) and \textit{hcp} ($AB$-$\dots$) stacking orders, respectively, as indicated in Fig.~\ref{fig:3}(d). 
Consequently, the initial $\gamma$/$\alpha$ interfacial transition region (cyan region in Figs.~\ref{fig:3}(b-d)) exhibits a stepped edge with two horizontally mismatched ($A’ \vert B$) and ($C’ \vert A$) O stacking layers and a vertically mismatched $B’$-$B$ stacking order which is energetically unfavorable. 
The dynamic evolution of the representative ($A’ \vert B$) O layers (shadowed layer in Fig.~\ref{fig:3}(d)) is further detailed in Fig.~\ref{fig:3}(e), and Supplementary Video I illustrates the complete simulation. 
Within the first 5 ps, the $\alpha$-O $B$ stacking layer reconstruct into a $\gamma$-O $C’$-like stacking, accompanied by overall vertical lattice distortion, leading to a transient local \textit{hcp}-like stacking. 
Further plane ``slip'' displacements follow typical hexagonal directions alone \hkl[110] or \hkl[1-10], as indicated by the blue arrows in Fig.~\ref{fig:3}(e). 
These plane displacements or ``glides'' complete at $t = 355$ ps, along with simultaneously rapid local Ga rearrangement (see Supplementary Video I, from frame 3450 to frame 3560, $t = 345\sim356$ ps). The final $\gamma$/$\alpha$ interface, presents a perfectly lattice-aligned O ($B’=B$) single layer with ultimate atomic sharpness.  

Despite that $\beta$-to-$\gamma$ phase transformations also occurs under ion irradiation, the mechanism of the disorder-induced transformations in $\alpha$-\ce{Ga2O3} is dramatically different from that in $\beta$-phase [15]. 
Indeed, previously it was demonstrated that $\beta$-to-$\gamma$ phase transformation occurs due to accumulation of Ga disorder, while O sublattice having \textit{fcc} structure for both phases exhibits a strong recrystallization trend within collision cascade [15,17]. 
In contrast, the O sublattice in $\alpha$-phase posses \textit{hcp} structure, so that the structural rearrangements in both sublattices are required for the $\alpha$-to-$\gamma$ phase transformation. 
Furthermore, out of the energy consideration, $\alpha$-to-$\gamma$ phase transformation requires an accumulation of the tensile strain in the system, see Supplementary Note I and Ref.~[30]. 
Literally, in order to realize the $\alpha$-to-$\gamma$ \ce{Ga2O3} phase transformation, the \textit{hcp} $\alpha$-O sublattice must transform to \textit{fcc} $\gamma$-O sublattice. 
Both \textit{hcp} and \textit{fcc} stackings share efficient closed-packing arrangements, suggesting that the transformation likely occurs via a slip of closed-packed layers. 
However, the atomic volume differences between two phases are the biggest among all polymorphs. 
The smallest atomic volume of the $\alpha$-phase is around 10.1 \r A$^{3}$ per atom while that of the $\gamma$-phase is around 11 \r A$^{3}$ per atom, see Supplementary Note 1. 
This indicates that such phase transformation needs to be assisted with an expansion of the system.

\begin{figure*}[ht!]
 \includegraphics[width=16cm]{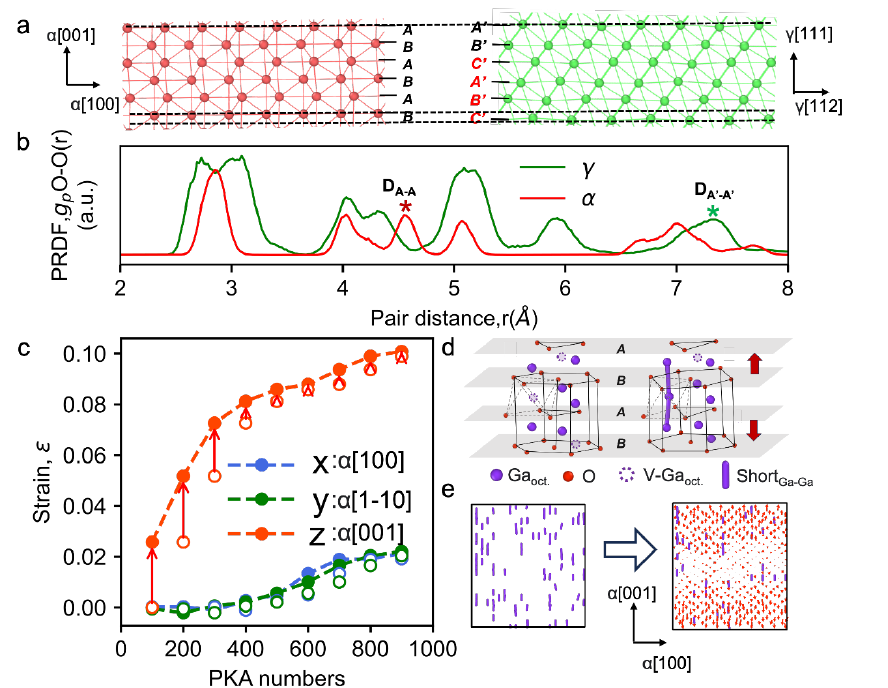}
 \caption{
    \textbf{Anisotropic expansion of $\alpha$-\ce{Ga2O3} triggered by disorder.} 
    (a) Oxygen sublattices of $\alpha$-\ce{Ga2O3} (left) and $\gamma$-\ce{Ga2O3} (right) shown as $A$-$B$ planes in the \textit{hcp} stacking for $\alpha$-\ce{Ga2O3} and as $A'$-$B'$-$C'$ planes in the \textit{fcc} stacking for $\gamma$-\ce{Ga2O3} stacked perpendicularly to the close-packed direction for both lattices. 
    (b) The corresponding partial radial pair distribution function (PRDF) curves for the oxygen slabs of $\alpha$-\ce{Ga2O3} (red) and $\gamma$-\ce{Ga2O3} (green) shown in (a). 
    Asterisks highlight the peaks of the $A$-to-$A$ and $A'$-to-$A'$ distances in both structures. 
    (c) The expansion fraction of the $\alpha$-\ce{Ga2O3} along the $x$, $y$, and $z$ axes during the collision cascade simulations. 
    The $x$, $y$, and $z$ axes correspond to $\alpha$\hkl[100] (blue), $\alpha$\hkl[1-10] (green), and $\alpha$\hkl[001] (red) orientations, respectively. 
    The hollow and solid circles correspond to the system before and after isothermal-isobaric ensemble ($NPT$) relaxation. 
    (d) Partial enlargement images of the pristine (left) and defective (right) $\alpha$-\ce{Ga2O3} during the collision cascade simulations. 
    The Ga and O atoms are large purple and small red spheres, respectively. 
    The short-distance Ga-Ga neighbors (distance shorter than 2.5 \r A) are highlighted with the purple sticks. 
    (e) The evolution of the number of short-distance Ga-Ga neighbors (purple sticks) randomly formed in $\alpha$-\ce{Ga2O3} after 100 PKA collision cascades from before (left) to after (right) $NPT$ relaxation. 
    The red arrows are the displacement vectors.
 }
 \label{fig:4}
\end{figure*}

To understand the expansion mechanism of $\alpha$-to-$\gamma$ phase transformation, we systematically compare O sublattice parameters of $\alpha$ and $\gamma$-\ce{Ga2O3} phases, as analyzed in Fig.~\ref{fig:4}. 
Indeed, Fig.~\ref{fig:4}(a) illustrates the stacking of the closed-packed planes perpendicular to the closed-packed layers of the $\alpha$-O sublattice ($AB$-$AB$-$AB$) and $\gamma$-O sublattice ($A’B’C’$-$A’B’C’$). 
These data indicate that the interlayer distances in the $\alpha$-O sublattice are smaller than those in the $\gamma$-O sublattice (see the corresponding levels marked by the black dashed lines in Fig.~\ref{fig:4}(a)). 
Further, the PRDF curves of the single layer in the O sublattice which is perpendicular to the close-packed layers (Fig.~\ref{fig:4}(a)), the $A$-$A$ distances in the $\alpha$-O sublattice is the peak of $D_{A\text{-}A}$ at 4.5 \r A, the $A’$-$A’$ distances in the $\gamma$-O sublattice is $D_{A’\text{-}A’}$ at 7.1 \r A. 
Accounting these values the average interlayer distances in the closed-packed layer of $\alpha$-O and $\gamma$-O sublattice is 2.25 \r A and 2.37 \r A, respectively. 
This indicates an expansion between the closed-packed layers of the $\alpha$-phase (alongside $\alpha$\hkl[001] orientations) during the phase transformation of $\sim5\%$. 
Within the closed-packed layers, as shown in Supplementary Note II, the PRDF curves indicate negligible expansion, particularly in longer distances.
This implies that even though there are some deviations between the arrangements of efficient packing, no prominent ``isotropic'' expansion is observed within these closed-packed planes. 

Meanwhile, anisotropic expansion of $\alpha$-\ce{Ga2O3} may play a prominent role in explaining the mechanism of $\alpha$ to $\gamma$-\ce{Ga2O3} phase transformation under irradiation. 
To investigate the expansion evolution, we employ additional isothermal-isobaric ensemble ($NPT$) relaxations after every 100 primary knock-on atoms (PKAs) overlapping collision cascade simulations to provide sufficient degree of freedom for the system volume adjustment. 
Fig.~\ref{fig:4}(c) displays the strain in the three coordination directions, revealing significant expansion along the $z$ axis, especially during the early stages of the radiation defects accumulation. 
Conversely, the system sizes in the $x$ and $y$ directions show minimal changes before 300 PKAs, followed by a slight expansion. 
This anisotropic expansion aligns with the lattice differences between $\alpha$ and $\gamma$-\ce{Ga2O3} phases. 
The lattice experiences the strongest expansion along the $z$ axis which is perpendicular to the closed-packed plane on the oxygen sublattice, while the $x$ and $y$ axes belong within the closed-packed plane and the expansion in these directions is very small. 
The structure in the left-hand side of Fig.~\ref{fig:4}(d) shows the lattice of the pristine $\alpha$-\ce{Ga2O3}. 
One can see that one of each three octahedral sites within the \textit{hcp} O sublattice is vacant. 
After collision cascades, the defective $\alpha$-\ce{Ga2O3} (the right-hand side of Fig.~\ref{fig:4}(d)) shows that the displaced Ga atoms readily occupy the available octahedral interstitial positions, forming Ga-Ga pairs with incredibly short distances of less than 2.5 \r A (visualized in Fig.~\ref{fig:4}(d) by purple sticks). 
Appearance of the multitude of the short-distance Ga-Ga pairs aligned with the $z$ axis in the defective $\alpha$-\ce{Ga2O3} (see Fig.~4(e)) demonstrates that Ga defects indeed primarily occupy available octahedral sites in the \textit{hcp} oxygen sublattice. 
These defects accumulate stress which relaxes straining the lattice in the $z$-axis direction as seen in experiment. 
Indeed, after the \textit{NPT} relaxation, the system expands along the $z$ axis, and the number of Ga-Ga short-distance pairs decreases significantly (see Fig.~\ref{fig:4}(e)). 
The preferential movement of atoms parallel to the $z$ axis can be seen via the displacement vectors for all the atoms in the lattice after relaxation, which are shown by the red arrows in the right-hand side of Fig.~\ref{fig:4}(e). 
In other words, our results suggest that the accumulation of Ga defects generates the stress between the oxygen closed-packed planes, leading to significant expansion. 
The fully relaxed stress in our simulations within the $NPT$ ensemble, the system expands by approximately 10\% in the $z$ axis. 
This expansion is greater than 5\% needed for transformation of the \textit{hcp} $\alpha$-\ce{Ga2O3} oxygen sublattice into the \textit{fcc} $\gamma$-\ce{Ga2O3} oxygen sublattice as according to the mechanism proposed in Fig.~\ref{fig:3}.
A full relaxation of the accumulated stress can be expected only near the surface. 
The large expansion of the lattice allows for easier accumulation of the defects, which leads to faster deterioration of the crystal lattice. 
This is why the lattice of $\alpha$-\ce{Ga2O3} near the surface does not transform into the stable $\gamma$-\ce{Ga2O3} phase under ion irradiation, but first becomes amorphous, see Fig.~\ref{fig:1}. 
However, in deeper regions, stress generated by octahedral Ga interstitials is not easily released, controlling interplane distances to be closer to the $\gamma$-O sublattice. 
As stress increases, crystal plane slip becomes highly likely, completing the phase transformation from $\alpha$ to $\gamma$ deep beneath the surface.

\section*{Conclusions}

Disorder-induced ordering and unprecedently high radiation tolerance in $\gamma$-phase of gallium oxide is a recent spectacular discovery at the intersection of the fundamental physics and electronic applications. 
Importantly, before the present work, all these amazing literature data were collected with initial samples in form of the thermodynamically stable $\beta$-phase of this material. 
Here, we investigated these phenomena starting instead from already metastable $\alpha$-phase and explained radically new trend occurring in the system. 
We argued that in contrast to that in $\beta$-to-$\gamma$ disorder-induced transitions, the O sublattice in $\alpha$-phase exhibits hexagonal close-packed structure, so that to activate $\alpha$-to-$\gamma$ transformation significant structural rearrangements are required in both Ga and O sublattices. 
Moreover, consistently with theoretical predictions, $\alpha$-to-$\gamma$ phase transformation requires accumulation of the substantial tensile strain to initiate otherwise impossible lattice glides. 
Thus, we explain the experimentally observed trends in term of the combination of disorder and strain governing the process. 
Finally, and perhaps most amazingly, we demonstrate atomically abrupt $\alpha$/$\gamma$ interfaces paradoxically self-organized out of the stochastic disorder.

\section*{Methods}

\subsection*{Experimental methods}

In the present work we used rhombohedral $\sim1$ \textmu m thick $\alpha$-\ce{Ga2O3} films grown on sapphire substrates by halide vapor phase epitaxy (see details of the synthesis elsewhere [31]). 
The samples were implanted at room temperature with 400 keV $^{58}$Ni$^{+}$ ions in a wide dose range ($1 \times 10^{15}$–$1 \times 10^{17}$ Ni/cm$^{2}$) keeping the ion flux constant at $6\times10^{12}$ atom$\cdot$cm$^{-2}\cdot$s$^{-1}$. 
All the implants were performed at $7\degree$ off-angle orientation from normal direction to minimize channeling. 
For each ion dose the corresponding displacements per atom (dpa) values were calculated using conventional methodology [32] based on SRIM code [28]simulations. 
Specifically, the quoted dpa values were taken at the maximum of the SRIM vacancy generation profiles simulated for a given ion dose and normalized to the atomic density of $\alpha$-\ce{Ga2O3} ($n_\mathrm{at} = 10.35 \times 10^{22}$ at./cm$^{3}$). 
The SRIM simulations were performed in a full damage cascade mode with 28 eV and 14 eV as the displacement energies for Ga and O atoms, respectively~[33].

Structural characterization of the implanted samples was performed by a combination of Rutherford backscattering spectrometry in channeling mode (RBS/C), x-ray diffraction (XRD), and scanning transmission electron microscopy (STEM). The RBS/C measurements were performed by 1.6 MeV He$^{+}$ ions incident along \hkl[001] direction in $\alpha$-\ce{Ga2O3} part of the structure and backscattered into a detector placed at $165\degree$ relative to the incident beam direction. 
XRD 2$\Theta$ measurements were performed using the RIGAKU SmartLab diffractometer with high-resolution Cu $K_{\alpha 1}$ radiation and Ge(440) four-bounced monochromator. 
For cross-sectional STEM studies, selected samples were thinned by mechanical polishing and by Ar ion milling in a Gatan PIPS II (Model 695), followed by plasma cleaning (Fishione Model 1020) immediately before loading the samples into a $C_{s}$-corrected Thermo Fisher Scientific Titan G2 60–300 kV microscope, operated at 300 kV. 
The STEM images were recorded using a probe convergence semi-angle of 23 mrad, a nominal camera length of 60 mm using two different detectors: high-angle annular dark field (HAADF) (collection angles 100–200 mrad), and bright field (BF) (collection angles 0–22 mrad). The structural model of different phases was displayed using VESTA software [34]

\subsection*{Computational methods}

The machine-learned molecular dynamics (ML-MD) simulations were conducted using LAMMPS package~[35]. 
The self-developed ML interatomic potential of \ce{Ga2O3} system was employed [30], which was designed with the high accuracy for all five experimentally known \ce{Ga2O3} polymorphs and generality for disordered structures. 
The evolution of $\gamma$/$\alpha$ interface are stimulated using an orthogonal cell comprising 15,360 atoms with the side lengths of $\sim82.2\times17.8\times113.4$ \r A$^{3}$. The $x$, $y$, and $z$ axes correspond to $\gamma$\hkl[112]/$\alpha$\hkl[100], $\gamma$\hkl[110]/$\alpha$\hkl[1-10], and $\gamma$\hkl[111]/$\alpha$\hkl[001] orientations, respectively. 
The cell is initially optimized to a local minimum and is further run at 900 K and 0 bar in isothermal-isobaric ensemble ($NPT$) using Nos{\'e}-Hoover algorithm [36] for 500 ps with 1 fs per MD step. 
The polyhedral template matching (PTM) method [37] is employed to identify the local stacking structure of O sublattice. 
The coordination number of Ga atoms are counted with the cutoff radius of 2.6 \r A. 
The structural analyses and visualization are done with OVITO software [38]. 

A total of 900 overlapping cascades were conducted on the $\alpha$-\ce{Ga2O3} cell, which contains 14,400 atoms in a $\sim51\times53\times55$ \r A$^{3}$ box. 
This scale of the simulation cell was taken to prevent cascade overlapping the temperature-controlled borders and to minimize computational time. 
In each iteration, Ga or O atom was randomly selected as the PKA. 
The PKA was assigned a kinetic energy of 500 eV with a uniformly random momentum direction. 
To maintain consistency, the entire cell was translated and wrapped at periodic boundaries, positioning the PKA at the center of the cell. 
Each simulation iteration consisted of two periods. 
During the first cascade period, the cell was thermalized using $NVE$-MD for 5,000 MD steps with an adaptive time step. 
Electron-stopping frictional forces were applied to atoms with kinetic energies above 10 eV [39,40]. 
In the subsequent period, the simulations continued in a quasi-canonical ensemble with a Langevin thermostat [41] applied to border atoms (within 7.5 \r A of the simulation box boundaries, redefined in each iteration) for 10 ps at 300 K. 
Additionally, relaxation simulations were periodically conducted after every 100 cascades. 
During these relaxation simulations, the system was subjected to isothermal-isobaric ensemble ($NPT$) conditions at 300 K and 0 bar for 100 ps, with temperature controlled by the Nos{\'e}-Hoover thermostat and barostat [36].

\section*{Acknowledgments}

M-ERA.NET Program is acknowledged for financial support via GOFIB project (administrated by the Research Council of Norway project number 337627 in Norway and the Academy of Finland project number 352518 in Finland). 
Additional support was received from the Research Council of Norway in the frame of the FRIPRO Program project number 739211. 
The experimental infrastructures were provided at the Norwegian Micro- and Nano-Fabrication Facility, NorFab, supported by the Research Council of Norway project number 295864, at the Norwegian Centre for Transmission Electron Microscopy, NORTEM, supported by the Research Council of Norway project number 197405. 
J.Z. acknowledges the National Natural Science Foundation of China under Grant 62304097; Guangdong Basic and Applied Basic Research Foundation under Grant 2023A1515012048; Shenzhen Fundamental Research Program under Grant JCYJ20230807093609019. 
Computing resources were provided by the Finnish IT Center for Science (CSC) and by the Center for Computational Science and Engineering at the Southern University of Science and Technology. 
The paper was also supported by the ``Strategic R\&D program'' funded by the Korea Institute of Ceramic Engineering and Technology (KICET), Republic of Korea, in 2024 (KPP23004-0-02). 
The international collaboration was also fertilized via INTPART Program at the Research Council of Norway project number 322382 as well as UTFORSK Program at the Norwegian Directorate for Higher Education and Skills project number UTF-2021/10210. 

\section*{Author contribution}

A.K. and A.A. conceived the research strategy and designed the methodological complementarities. 
J.H.P. and D.W.J. contributed to the crystal growth and provided the samples. 
A.A. and J.G.F. carried out experiments and provided initial drafts for the description of the experimental data. 
R.H. and J.Z performed molecular dynamics simulations. 
F.D., R.H. and J.Z. developed the theoretical models and composed the theoretical part of the manuscript. 
A.K. and A.A. finalized the manuscript with the input from all the co-authors. 
All co-authors discussed the results as well as reviewed and approved the manuscript. 
A.K., {\O}.P., J.Z., and F.D, administrated their parts of the project and contributed to the funding acquisition. 
A.K. coordinated the work of the partners.

\section*{Competing interests} 

The authors declare no competing interests.

\section*{References}

\noindent$[1]$ I. Cora, Zs. Fogarassy, R. Fornari, M. Bosi, A. Re{\v c}nik, and B. P{\'e}cz, 
``In situ TEM study of $\kappa \rightarrow \beta$ and $\kappa \rightarrow \gamma$ phase transformations in Ga$_2$O$_3$'', 
\textit{Acta Mater.}, \textbf{183}, 216--227 (2020).

\noindent$[2]$ C. Wouters, M. Nofal, P. Mazzolini, J. Zhang, T. Remmele, A. Kwasniewski, O. Bierwagen, and M. Albrecht, 
``Unraveling the atomic mechanism of the disorder–order phase transition from $\gamma$-Ga$_2$O$_3$ to $\beta$-Ga$_2$O$_3$'', 
\textit{APL Mater.}, \textbf{12}, 011110 (2024).

\noindent$[3]$ J. Wang, X. Guan, H. Zheng, L. Zhao, R. Jiang, P. Zhao, Y. Zhang, J. Hu, P. Li, S. Jia, and J. Wang, 
``Size-dependent phase transition in ultrathin Ga$_2$O$_3$ nanowires'', 
\textit{Nano Lett.}, \textbf{23}, 7364--7370 (2023).

\noindent$[4]$ S.-C. Zhu, S.-H. Guan, and Z.-P. Liu, 
``Mechanism and microstructures in Ga$_2$O$_3$ pseudomartensitic solid phase transition'', 
\textit{Phys. Chem. Chem. Phys.}, \textbf{18}, 18563 (2016).

\noindent$[5]$ K. R. Gann, C. S. Chang, M.-C. Chang, D. R. Sutherland, A. B. Connolly, D. A. Muller, R. B. van Dover, and M. O. Thompson, 
``Initial nucleation of metastable $\gamma$-Ga$_2$O$_3$ during sub-millisecond thermal anneals of amorphous Ga$_2$O$_3$'', 
\textit{Appl. Phys. Lett.}, \textbf{121}, 062102 (2022).

\noindent$[6]$ X. Chen, F. Ren, S. Gu, and J. Ye, 
``Review of gallium-oxide-based solar-blind ultraviolet photodetectors'', 
\textit{Photonics Res.}, \textbf{7}, 381--415 (2019).

\noindent$[7]$ S. J. Pearton, J. Yang, P. H. Cary IV, F. Ren, J. Kim, M. J. Tadjer, and M. A. Mastro, 
``A review of Ga$_2$O$_3$ materials, processing, and devices'', 
\textit{Appl. Phys. Rev.}, \textbf{5}, 011301 (2018).

\noindent$[8]$ A. J. Green, J. Speck, G. Xing, P. Moens, F. Allerstam, K. Gumaelius, T. Neyer, A. Arias-Purdue, V. Mehrotra, A. Kuramata, K. Sasaki, S. Watanabe, K. Koshi, J. Blevins, O. Bierwagen, S. Krishnamoorthy, K. Leedy, A. R. Arehart, A. T. Neal, S. Mou, S. A. Ringel, A. Kumar, A. Sharma, K. Ghosh, U. Singisetti, W. Li, K. Chabak, K. Liddy, A. Islam, S. Rajan, S. Graham, S. Choi, Z. Cheng, and M. Higashiwaki, 
``$\beta$-Gallium oxide power electronics'', 
\textit{APL Mater.}, \textbf{10}, 029201 (2022).

\noindent$[9]$ R. Zhu, H. Liang, S. Liu, Y. Yuan, X. Wang, F. C.-C. Ling, A. Kuznetsov, G. Zhang, and Z. Mei, 
``Non-volatile optoelectronic memory based on a photosensitive dielectric'', 
\textit{Nat. Commun.}, \textbf{14}, 5396 (2023).

\noindent$[10]$ M. J. Tadjer, 
``Toward gallium oxide power electronics'', 
\textit{Science}, \textbf{378}, 724 (2022).

\noindent$[11]$ A. Azarov, C. Bazioti, V. Venkatachalapathy, P. Vajeeston, E. Monakhov, and A. Kuznetsov, 
``Disorder-induced ordering in gallium oxide polymorphs'', 
\textit{Phys. Rev. Lett.}, \textbf{128}, 015704 (2022).

\noindent$[12]$ H.-L. Huang, C. Chae, J. M. Johnson, A. Senckowski, S. Sharma, U. Singisetti, M. H. Wong, and J. Hwang, 
``Atomic scale defect formation and phase transformation in Si implanted $\beta$-Ga$_2$O$_3$'', 
\textit{APL Mater.}, \textbf{11}, 061113 (2023).

\noindent$[13]$ T. Yoo, X. Xia, F. Ren, A. Jacobs, M. J. Tadjer, S. Pearton, and H. Kim, 
``Atomic-scale characterization of structural damage and recovery in Sn ion-implanted $\beta$-Ga$_2$O$_3$'', 
\textit{Appl. Phys. Lett.}, \textbf{121}, 072111 (2022).

\noindent$[14]$ E. A. Anber, D. Foley, A. C. Lang, J. Nathaniel, J. L. Hart, M. J. Tadjer, K. D. Hobart, S. Pearton, and M. L. Taheri, 
``Structural transition and recovery of Ge implanted $\beta$-Ga$_2$O$_3$'', 
\textit{Appl. Phys. Lett.}, \textbf{117}, 152101 (2020).

\noindent$[15]$ A. Azarov, J. García Fernández, J. Zhao, F. Djurabekova, H. He, R. He, Ø. Prytz, L. Vines, U. Bektas, P. Chekhonin, N. Klingner, G. Hlawacek, and A. Kuznetsov, 
``Universal radiation tolerant semiconductor,'' 
\textit{Nat. Commun.}, \textbf{14}, 4855 (2023).

\noindent$[16]$ A. Y. Polyakov, A. A. Vasilev, A. I. Kochkova, I. V. Shchemerov, E. B. Yakimov, A. V. Miakonkikh, A. V. Chernykh, P. B. Lagov, Y. S. Pavlov, A. S. Doroshkevich, R. Sh. Isaev, A. A. Romanov, L. A. Alexanyan, N. Matros, A. Azarov, A. Kuznetsov, and S. Pearton, 
``Proton damage effects in double polymorph $\gamma/\beta$-Ga$_2$O$_3$ diodes'', 
\textit{J. Mater. Chem. C}, \textbf{12}, 1020 (2024).

\noindent$[17]$ J. Zhao, J. Garc{\'i}a-Fern{\'a}ndez, A. Azarov, R. He, {\O}. Prytz, K. Nordlund, M. Hua, F. Djurabekova, and A. Kuznetsov, 
``Crystallization instead amorphization in collision cascades in gallium oxide'', 
\textit{arXiv}, 2401.07675 (2023).

\noindent$[18]$ R. He, J. Zhao, J. Byggm{\"a}star, H. He, and F. Djurabekova, 
``Ultrahigh Stability of O-Sublattice in $\beta$-Ga$_2$O$_3$'', 
\textit{arXiv}, 2404.10451 (2024).

\noindent$[19]$ A. Azarov, C. Radu, A. Galeckas, I. F. Mercioniu, A. Cernescu, V. Venkatachalapathy, E. Monakhov, F. Djurabekova, C. Ghica, J. Zhao, and A. Kuznetsov, 
``Self-assembling of multilayered polymorphs with ion beams'', 
\textit{arXiv}, 2404.19572 (2024).

\noindent$[20]$ A. I. Titov, K. V. Karabeshkin, A. I. Struchkov, V. I. Nikolaev, A. Azarov, D. S. Gogova, and P. A. Karaseov, 
``Comparative study of radiation tolerance of GaN and Ga$_2$O$_3$ polymorphs'', 
\textit{Vacuum}, \textbf{200}, 111005 (2022).

\noindent$[21]$ A. Azarov, J.-H. Park, D.-W. Jeon, and A. Kuznetsov, 
``High mobility of intrinsic defects in a-Ga$_2$O$_3$'', 
\textit{Appl. Phys. Lett.}, \textbf{122}, 182104 (2023).

\noindent$[22]$ D. Tetelbaum, A. Nikolskaya, D. Korolev, T. Mullagaliev, A. Belov, V. Trushin, Y. Dudin, A. Nezhdanov, A. Mashin, A. Mikhaylov, A. Pechnikov, M. Scheglov, V. Nikolaev, and D. Gogova, 
``Ion-beam modification of metastable gallium oxide polymorphs'', 
\textit{Mater. Lett.}, \textbf{302}, 130346 (2021).

\noindent$[23]$ S. O. Kucheyev, J. S. Williams, C. Jagadish, J. Zou, and G. Li, 
``Damage buildup in GaN under ion bombardment'', 
\textit{Phys. Rev. B}, \textbf{62}, 7510 (2000).

\noindent$[24]$ K. Lorenz, E. Wendler, A. Redondo-Cubero, N. Catarino, M.-P. Chauvat, S. Schwaiger, F. Scholz, E. Alves, and P. Ruterana, 
``Implantation damage formation in a-, c- and m-plane GaN'', 
\textit{Acta Mater.}, \textbf{123}, 177--187 (2017).

\noindent$[25]$ S. Fujita, M. Oda, K. Kaneko, and T. Hitora, 
``Evolution of corundum-structured III-oxide semiconductors: Growth, properties, and devices'', 
\textit{Jpn. J. Appl. Phys.}, \textbf{55}, 1202A3 (2016).

\noindent$[26]$ A. Galeckas et al., 
``Optical library of Ga$_2$O$_3$ polymorphs'', under preparation.

\noindent$[27]$ C. Wu, C. He, D. Guo, F. Zhang, P. Li, S. Wang, A. Liu, F. Wu, and W. Tang, 
``Vertical $\alpha/\beta$-Ga$_2$O$_3$ phase junction nanorods array with graphene-silver nanowire hybrid conductive electrode for high-performance self-powered solar-blind photodetectors'', 
\textit{Mat. Today Phys.}, \textbf{12}, 100193 (2020).

\noindent$[28]$ J. F. Ziegler, M. D. Ziegler, and J. P. Biersack, 
``SRIM—the stopping and range of ions in matter (2010)'', 
\textit{Nucl. Instrum. Methods Phys. Res., Sect. B}, \textbf{268}, 1818 (2010).

\noindent$[29]$ O. Nikulina, D. Yatsenko, O. Bulavchenko, G. Zenkovets, and S. Tsybulya, 
``Debye function analysis of nanocrystalline gallium oxide $\gamma$-Ga$_2$O$_3$'', 
\textit{Z. Kristallogr.}, \textbf{231}, 261--266 (2016).

\noindent$[30]$ J. Zhao, J. Byggm{\"a}star, H. He, K. Nordlund, F. Djurabekova, and M. Hua, 
``Complex Ga$_2$O$_3$ polymorphs explored by accurate and general-purpose machine-learning interatomic potentials'', 
\textit{npj Comput. Mater.}, \textbf{9}, 159 (2023).

\noindent$[31]$ H. Son and D.-W. Jeon, 
``Optimization of the growth temperature of $\alpha$-Ga$_2$O$_3$ epilayers grown by halide vapor phase epitaxy'', 
\textit{J. Alloys Compd.}, \textbf{773}, 631 (2019).

\noindent$[32]$ A. I. Titov, A. Yu. Azarov, L. M. Nikulina, and S. O. Kucheyev, 
``Damage buildup and the molecular effect in Si bombarded with PFn cluster ions'', 
\textit{Nucl. Instrum. Methods Phys. Res., Sect. B}, \textbf{256}, 207 (2007).

\noindent$[33]$ B. R. Tuttle, N. J. Karom, A. O’Hara, R. D. Schrimpf, and S. T. Pantelides, 
``Atomic-displacement threshold energies and defect generation in irradiated $\beta$-Ga$_2$O$_3$: A first-principles investigation'', 
\textit{J. Appl. Phys.}, \textbf{133}, 015703 (2023).

\noindent$[34]$ K. Momma and F. Izumi, 
``VESTA 3 for three-dimensional visualization of crystal, volumetric and morphology data'', 
\textit{J. Appl. Crystallogr.}, \textbf{44}, 1272--1276 (2011).

\noindent$[35]$ A. P. Thompson, H. M. Aktulga, R. Berger, D. S. Bolintineanu, W. M. Brown, P. S. Crozier, P. J. in’t Veld, A. Kohlmeyer, S. G. Moore, T. D. Nguyen, R. Shan, M. J. Stevens, J. Tranchida, C. Trott, and S. J. Plimpton, 
``LAMMPS - a flexible simulation tool for particle-based materials modeling at the atomic, meso, and continuum scales'', 
\textit{Comput. Phys. Commun.}, \textbf{271}, 108171 (2022).

\noindent$[36]$ W. G. Hoover, 
``Canonical dynamics: Equilibrium phase-space distributions'', 
\textit{Phys. Rev. A}, \textbf{31}, 1695 (1985).

\noindent$[37]$ P. M. Larsen, S. Schmidt, and J. Schi{\o}tz, 
``Robust structural identification via polyhedral template matching'', 
\textit{Modell. Simul. Mater. Sci. Eng.}, \textbf{24}, 055007 (2016).

\noindent$[38]$ A. Stukowski, 
``Visualization and analysis of atomistic simulation data with OVITO – the open visualization tool'', 
\textit{Modell. Simul. Mater. Sci. Eng.}, \textbf{18}, 015012 (2010).

\noindent$[39]$ K. Nordlund, 
``Molecular dynamics simulation of ion ranges in the 1–100 keV energy range'', 
\textit{Comput. Mater. Sci.}, \textbf{3}, 448 (1995).

\noindent$[40]$ K. Nordlund, M. Ghaly, R. S. Averback, M. Caturla, T. Diaz de la Rubia, and J. Tarus, 
``Defect production in collision cascades in elemental semiconductors and fcc metals'', 
\textit{Phys. Rev. B}, \textbf{57}, 7556 (1998).

\noindent$[41]$ B. D{\"u}nweg and W. Paul, 
``Brownian dynamics simulations without gaussian random numbers'', 
\textit{Int. J. Mod. Phys. C}, \textbf{02}, 817 (1991).




\end{document}